Security and Privacy in Decentralized Networks

# SAHER: Secure and Efficient Routing in Sensor Networks

Minko Dudev[*]  Sebastian Gerling[†]  Philip Peter[‡]

March 2007

___________________________

[*]mdudev@mpi-inf.mpg.de

[†]sgerling@mpi-inf.mpg.de

[‡]philip.peter@gmx.de

# Contents

# 1 Abstract

As an increasing amount of research is being done on various applications of sensor networks in adversarial environments, ensuring secure routing becomes of critical importance for the success of such deployments. The problem of designing a secure routing protocol for ad hoc networks has been already addressed, yet, there exists no complete solution that meets the specific requirements of sensor networks, where nodes are extremely constrained in terms of both power and computational resources. Thus, we propose a new protocol that is not built solely around security but also has efficiency and simplicity among its main goals. We propose the Secure Ad Hoc Efficient Routing protocol (SAHER) which employs a two-tier architecture based on node clustering. Also, we combine mechanisms like local-scale geographic routing, per-node reputation tables, credit based alternate route enforcement and cumulative authentication. Using these techniques we examine ways to efficiently defend against the two most common network layer attacks: selective packet dropping and message flooding. Further, we consider join/leave operations which have not yet been studied in sufficient depth for sensor networks from a security standpoint. We provide a description of the protocol along with comprehensive experimental evaluation under different node distributions, different proportions of non-malicious vs. malicious nodes and different types of activity that malicious nodes could exhibit.

# 2 Introduction

In recent years Wireless Sensor Networks (WSNs) have been gaining considerable popularity among researchers as new applications for them are being explored. WSNs share a lot of commonality with the broader class of wireless ad-hoc networks, just like Mobile Ad-hoc Networks (MANets, e.g. JTRS [24]) or similar wireless mesh networks. A sensor network is set up by using many, small and cheap wireless nodes that are spread over an area from which they gather information. The nodes could have sensors for humidity or temperature as well as sensors to measure acceleration and motion, sensors to detect light or even microphones for sensing sounds. Interestingly, sensor nodes have also been proposed for the collection of biomedical data [16]. The use of WSNs can be imagined in a variety of applications like the surveillance of infrastructure (e. g. intrusion detection) as well as health-care or military applications. For the military such networks could be exploited as part of command and control, communications, reconnaissance and targeting systems (C4ISRT). There are already several applications where sensor networks have been extensively evaluated:

- Great Duck Island: In-Situ Sensor Networks for Habitat Monitoring [17]
- ZebraNet: A Wildlife Tracker [18]
- Meteorology and Hydrology in Yosemite National Park [19]
- Code Blue: Wireless Sensor Networks for Medical Care, e. g. a wireless pulse oximeter and a 2-lead EKG [20]
- PinPtr: Shooter Localization and Weapon Classification with Soldier-Wearable Networked Sensors [21]
- inTrack: High Precision Tracking of Mobile Sensor Nodes [22]
- Boomerang: Shooter detection, locating the attacker [23]

Some of the above and many other potential applications are mission critical and security is of prime importance. Therefore sensor nodes have to be designed with security in mind on all levels. From tamper-proof hardware all the way to security on the application layer. An area where a sophisticated attacker could mount an attack is at the network layer. Since most of the existing routing protocols for sensor networks were not designed with the goal to provide security, we introduce a protocol that is secure against the most common types of attacks. The protocol is coupled with a clustered architecture which promotes energy efficiency and scalability. When designing protocols for operation in adversarial environments many things have to be considered. Nodes could simply get captured, malicious nodes could flood the network with messages or just drop messages they are supposed to route. Beyond these primitive malicious node actions, a more intricate malicious node behavior could involve a combination of selective data packet dropping, ACK dropping, message flooding and providing false routing information. To detect malicious nodes and to defend against the disruptions they try to introduce is a difficult task. Explicit detection and exclusion of malicious nodes might not even be possible in completely decentralized environments. That is why we approach the problem by initially trusting all nodes in the network and gradually adapting to the network conditions to avoid routing to nodes that are deemed insecure. Decisions on how reliable a neighbor node is are made locally, by every node on the path. As in other secure routing protocols proposals the goal is to ensure that the network is still functional even if there are many malicious nodes present.

Examples of particular sensor node designs are the Berkeley Mica Motes [2], the PicoRadio Nodes [3], the EYES project [11] or the TMote [12]. A Mica Mote sensor node measures 5.7 x 3.18 x0.64 centimeters and consists of five major parts: a processing part containing the main microcontroller, a radio part, a part responsible for the power management, an I/O expansion and a secondary storage part. Compared to any modern day general-purpose computer all of them are extremely constrained. A node like this is equipped with a 4MHz processor, 128KB flash memory, 4K SRAM and has a radio range of around 30 meters. Nodes are also designed to work with very cheap, low capacity energy sources. A typical design could include a pair of AA batteries. On the other hand, in most real-world deployments prolonged operation is highly desirable. It would be very impractical to replace the batteries of already deployed nodes. In the worst case physical access to nodes might not even be possible. Thus, efficient operation is a must in WSN environments. Typical low level power management techniques include alternating sleep/listen cycles to save resources whenever possible. However, the amount of idle time for nodes would be highly dependent on the particular application. Here we focus on efficiency at a higher level, namely, routing. Unfortunately security and efficiency do not usually go hand-in-hand. Security mechanisms often are quite expensive in terms of computational power and design complexity. That is why we strive for a good balance with the architectural design and the routing techniques described in this report.

In environments where nodes are randomly distributed and are made of the same hardware the ability to self organize in a completely decentralized multi-hop ad-hoc network is of prime importance. While base stations could simplify the structure of the network, in practice, it might not be possible to deploy base stations. Even if they are available they are likely to become easy targets of simple, yet powerful physical layer attacks.

Secure routing protocols designed for traditional computers are generally computationally expensive or are not applicable for a fully decentralized environment. Sensor nodes have limited energy resources, limited computational resources and are often expected to be fully self organizing. The design of a WSN relies on the assumption that nodes are cheap so that the coverage of large areas in likely hostile environments, where retrieving the nodes might not be possible, is feasible. We need a secure routing protocol that deals with these constraints. Some research has already been done on the design of secure sensor networks, mostly applying vari-ants of known techniques from traditional networking. Security can be added at different layers depending on the expected adversarial behavior against which the design has to defend. There has been research focused on applying security methods like end-to-end encryption using AES [14] or Sensor Protocols for Information via Negotiation (SPINs) [13] to prevent malicious nodes from eavesdropping on the communication channel, the introduction of new malicious nodes, corrupting messages, dropping messages or Denial of Service Attacks. Perrig et al. introduced SPINS[13] which consists of two protocols: the Secure Network Encryption Protocol (SNEP) and the $\mu$TESLA, a "micro"version of the Timed, Efficient, Streaming, Loss-tolerant Authentication Protocol [15]. SNEP provides authentication, confidentiality and data freshness; $\mu$TESLA is responsible for authenticated broadcast. There are also some key distribution schemes developed specifically for resource constrained ad hoc environments, for example, PIKE [1] providing for the setup of secure node-to-node connection through the establishment of pairwise shared keys stored at specific intermediate peers. There are also Geographic Hash Tables (GHTs) that introduce redundancy in the network by storing replicas of data on several objects for fault tolerance.

Additional complexity is added by the fact that the network might not be static. Some applications could require nodes to leave and join the network or change their locations periodically. Yet, very few proposals discuss join/leave operations because it is challenging to provide authentication and to detect when nodes are faking their identities in a resource efficient way.

Due to the limitations mentioned at the beginning, it is necessary to keep the routing tables as small as possible. This reduces the needed per node storage as well as the amount of routing table updates which traverse the network since nodes do not need to have perfect knowledge of the whole network. Not maintaining enough routing information however leads often to the situation where data messages have to be broadcasted. Routing in ad-hoc networks where a positioning system is available could be implemented efficiently using a variant of geographic routing. In geographic routing the main principle is not to use logical but physical addresses for the destination, usually represented as global location coordinates. This lead, for example, to the Greedy Perimeter Stateless Routing in Wireless Networks (GPSR) protocol [6]. In this class of protocols the message gets routed to a node that is the geographically nearest to the given destination position. This is a highly efficient routing method which could help the network scale to many thousands of nodes as an optimal or nearly optimal path will be chosen every time. However generic geographic protocols have not been designed with security among their goals. Interesting proposals for secure routing in sensor networks include Secure Cell Relay (SCR) routing protocol [4] and the Secure Sensor Network Routing (SSNR) [5]. They introduce security but still appear to suffer from inefficiency and complexity.

Besides the securing of routing protocols for WSNs there are a lot of further security issues that are of interest. The simplest yet extremely efficient attack is spread-spectrum radio jamming. An attacker with virtually unlimited resources could introduce noise at all frequencies over the area where the sensor network is deployed which prevents any message exchange. It is practically impossible to defend against this type of attacker unless alternative communications channels exit. However if the attacker is not omnipresent or does not have unlimited power – coding, frequency hopping and randomized send/listen/sleep cycles could be effective counter measures. Other challenges for the security of WSNs at higher levels include securing the key exchange of pairwise keys between nodes without having a base station and the secure location verification of nodes.

Our approach is held as simple as possible while still achieving enough security to get messages routed successfully or at least with very high probability when there are nodes that drop packages on the path. We have aimed at minimal overhead when the network is in normal operation and malicious nodes are not present. We cluster the whole network and introduce a two-tier architecture. In one cluster every node knows the location and identifier of any other member of the same cluster. Inside each cluster geographic routing is used. This approach of separating the network into neighborhoods allows it to scale to arbitrary size because nodes only need to keep information about the clusters they belong to in their routing tables. Every cluster has border nodes which are highly connected to border nodes of other clusters. Border nodes handle communication between clusters. They do not know what exactly the topology of the foreign cluster they are connected to is, but they know what node ids belong to it. Internal nodes and border nodes are physically not different types of nodes, which means that every node can become a border node when it fulfills the connectivity requirements. We use cumulative authentication [10] between nodes to implement ACKs that include route information efficiently. The protocol is based on avoiding high-risk paths rather than explicitly excluding the malicious nodes from the network. The initialization phase in which the nodes cluster together can be expensive, therefore if nodes are manually placed, clusters can be preconfigured or nodes with more power can be assigned as cluster heads (supernodes). Here we only explore the case where nodes are randomly distributed and physically identical.

The rest of this report is organized as follows. In the next section we describe in detail which assumptions underlay the design of our routing protocol. Also, we describe the scope of this study. Afterwards, in section 4 we explain the design of SAHER in detail. The design part of our report delves into the clustering of the nodes into groups, border node selection and then message routing itself. The design part is followed by a section on the security and efficiency analysis as measured by our simulator. Finally we conclude by summarizing the ideas introduced here and presenting future research directions.

# 3  Assumptions and Scope

We make a minimal set of assumptions that are enough to generalize our approach to most sensor network environments. In this work we make no assumptions that limit the design of the protocol to specific hardware. Nodes however must have a way to route geographically with a variant of GPSR[6] either by using full-fledged GPS receivers to determine their global coordinates, or by means of relative coordinates computation performed independently by the nodes in the initialization stage of the protocol. Rao et al. in [7] have suggested a way to mimic geographic routing by employing DHT functionality without the need of actual location information. Although this solution is interesting it should be noted that running a DHT over sensor nodes may not be feasible at this point and adding security mechanisms to the overlay will require even more resources if such mechanisms are at all possible [8]. We do not delve further into how exactly coordinates are determined, stored or retrieved. We do not assume the presence of a base station or strategically placed nodes with better hardware capabilities, which are acting as super-nodes to coordinate routing. An assumption for SAHER is that in the initialization stage each node can establish a pairwise shared key with any other node. This can be achieved in an elegant and efficient way by using the PIKE scheme [1], which offers a reasonable level of security in the presence of compromised nodes while maintaining very high memory efficiency. It should be emphasized that the usage of symmetric cryptographic techniques is crucial for the design of efficient protocols. Since it is widely known that standard asymmetric cryptography can be several orders of magnitude slower. Thus, asymmetric cryptographic primitives are not realistic when used as part of protocols related to current sensor node designs. It is reasonable to assume that in geographic routing nodes must have different coordinates thus preventing Sybil attacks are not a concern. For the correct operation of routing protocols it is natural to assume that ids are unique. A distributed way to detect replicas is presented in [9]. SAHER does not assume a static network layout, instead nodes can leave or join the network at any time.

The scope of this report mostly encompasses research into the security benefits of the two main ideas behind the protocol: node clustering and per-node risk-driven routing decisions. Here we do not focus on attacks that could occur outside the area of routing. We assume that nodes which are within range can successfully receive each other's frames because jamming does not occur or there are techniques in place that successfully defend against it.

# 4 Protocol Design

The main idea behind our design is the principle of isolation. That is, a primary goal in designing SAHER was to avoid global decisions. For this reason we implement isolation on two levels. First, nodes form clusters that guarantee that routing decisions have only local impact - nodes are oblivious to how messages are routed in clusters other than their own. Secondly, inside a cluster each node makes decisions where to forward a message to next, based only on its own observations of the reliability of its neighbors. Nodes are also dynamically separated into two classes: border nodes and internal nodes. Border nodes constitute nodes that are on the periphery of a cluster and dedicate their resources exclusively to forwarding messages between clusters. Internal nodes handle the main tasks for which the sensor network is deployed and participate in the message forwarding inside a cluster.

This basic design has manifold advantages. Geographic routing without a centralized lookup table of locations for each id would require $O(n)$ space in each node for $n$ nodes deployed in the network. Additionally, the whole network has to be flooded if a node changes location. Within a group however, nodes only need to keep track of the ids of other nodes in the group. Whole clusters of nodes can be isolated from routing if the cluster is deemed unreliable (malicious). In practice it is actually likely that malicious nodes will be introduced so that they are geographically close. Because of the design of the protocol in the initialization step (where clusters are formed), newly introduced malicious nodes are very probable to be forced into a cluster of their own. Fair nodes avoid talking to unreliable neighbors rather than punishing them in any active way. Decisions that help avoiding unsecure paths are made on a per-node basis. Each node can decide which of its neighbors should be trusted to forward the message further. Avoiding forwarding through malicious nodes is slightly different within a cluster and in between clusters, the details are described in the sections that follow.

## 4.1 Initialization

Upon deployment the nodes should self-organize into clusters and determine border nodes for each cluster. Nodes within each cluster must have perfect knowledge of each other's coordinates and ids. Nodes should share keys with their neighbors and should be able to acquire new pairwise shared keys on demand. This can be achieved by running a suitable variation (2D or 3D) of the PIKE protocol [1]. Initially, before clusters are formed all communication is achieved by flooding. The initialization is expected to be the most resource intensive stage of the protocol.

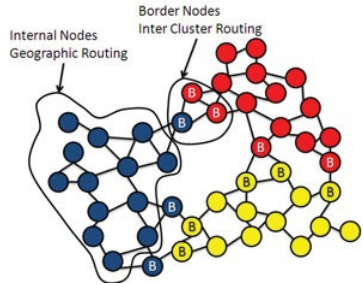
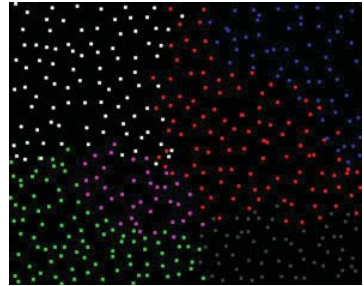

Figure 1: Design overview: different clusters and their border nodes, marked with 'B'.

Figure 2: Sample clustering result from the SAHER simulator.

#### 4.1.1 Cluster Formation

The formation of clusters has a large influence on the operation of SAHER at later stages, that is why it is crucial that clusters are formed in such a way that their internal connectivity and their connectivity to other clusters facilitates the optimal operation of the protocol. A desirable property is that clusters are sized in such a way that there are enough nodes per cluster so that the benefits of geographic routing make a difference. It is also desirable that nodes inside a cluster are well connected for several alternative routes to exist. These could be used for routing in the case of voids, malicious nodes or enforced alternative routing. Too large clusters can diminish the benefits of isolation. On the other extreme, not only geographic routing will stop making sense but also a large number of nodes will become border nodes - unable to perform the main tasks for which the sensor network is deployed. Optimally, there will be more than one border node connecting each cluster to any other neighboring cluster, to make sure clusters don't get overloaded and that a single failure in a connecting cluster will not disconnect a cluster from the rest of the network. Unfortunately making sure that these properties exist for clustering over every node distribution is a complex task that requires global knowledge of the network. There have been several distributed algorithms that provide efficient clustering (e.g. VCA [26]). However, to the best of our knowledge they are not secure. That is why we describe some simple approaches to clustering that perform reasonably well. We suggest three mechanisms that produce a reasonably good clustering and are not too resource demanding.

- The simplest way to split the network into clusters is to predefine a maximum hop radius for each cluster and let random nodes become cluster initiators. The cluster initiator would send out a message that floods its neighborhood within some hop count. If nodes already belong to a cluster they do not respond. Otherwise they join the cluster by adapting the node id of the initiator node as their group id. An undesirable situation where too many small clusters are formed might arise if several nodes in immediate proximity send a cluster initialization message at the same time. The message could traverse just a few nodes (in the worst case, none) before it reaches nodes that already belong to a different cluster. To alleviate this problem, nodes can be preconfigured to wait for a certain amount of time before they send out a cluster initialization message. For example, the time to wait can be a function of the node id so that all nodes have specific time at which they can attempt to create a new cluster. Of course, if a node already belongs to a cluster, it will not try to create a new one. Malicious nodes can be prevented from altering the traversed hop count with techniques like hash chain trees as described in [10]. This clustering approach works well in networks with randomly distributed nodes and allows limiting the size of the cluster to a certain diameter.

- A way to limit the number of nodes in a cluster to a particular number is to perform the random cluster formation procedure as described above but to allow only a certain number of nodes to join. This can be done by letting new nodes to register with all other nodes already in the cluster after receiving the invitation message for the cluster. Each node keeps track of how many nodes are registered in the group already. If a node wants to join, but the limit for the cluster has been exceeded it is notified and excluded from the forwarding tables of fair nodes in this cluster.

- Alternatively, we can approximate actual clusters based on physical proximity in non-uniformly distributed networks by allowing neighboring nodes to exchange their neighbor lists. If two neighboring nodes share a certain number of neighbors they can form a cluster. If a node is in between two clusters and shares the minimum number of common nodes with both clusters it will naturally select to be a part of the group with which it is better connected. In practice it may be useful to actually exchange neighbor lists within two or more hops. This and the shared number of neighbors threshold depends on how dense the individual clusters are expected to be. The authenticity of the neighbor lists can be established by adding a Message Authentication Code (MAC) to every neighbor discovery message and response. These MAC signed messages are then exchanged. This approach does not provide a limit on the number of nodes that can be part of the same cluster but creates very well connected clusters.

### 4.1.2 Border Node Selection

Border nodes are the interface through which each cluster communicates with the rest of the network. Border nodes dedicate all their resources to this task. To limit the impact that a malicious border node can have, a packet originating from a border node will be dropped by any non-malicious node that receives it. On one hand it is important to have enough border nodes to guarantee the existence of alternative routes; on the other hand it is highly undesirable to have a situation in which we have more border nodes than internal nodes. A simple approach is to let every node that has nodes from clusters other than its own within range, to be a border node. This method is simple and the problems mentioned above are taken care of simply if the clustering algorithm produces large-enough clusters. In a deployment where the formation of small clusters with many border nodes is inevitable there is a variety of ways to limit their number:

- Set hard thresholds on the minimum number of connections a border node should have with the internal nodes and the number of nodes from the neighboring cluster within range.

- Each node that has connections to outside clusters has a predefined probability of becoming a border node.

- In a similar manner to one of the ways to establish clustering, border nodes that connect to the same clusters exchange neighbor lists. A number of nodes that have the least amount of neighbors convert to internal nodes.

## 4.2 Routing Messages

Once the initialization is complete, routine routing can take place. Inside a cluster, nodes route as they would do in standard geographic routing protocols. When a destination node id is not found in the tables for the cluster messages will be forwarded to the border nodes. Border nodes on the other hand maintain some information about what node ids can be found in other clusters.

Internal nodes judge the reliability of border nodes and internal nodes in similar ways: they maintain a risk score for every immediate neighbor and every border node for the cluster. Then the routing decision is based on the minimization of a risk function. The risk function can be arbitrarily complex depending on how many parameters nodes keep tracks of from their neighbors. For routing inside a cluster we propose the simple function: $r = \frac{d}{t}$, where $r$ (risk) is the risk which we want to minimize, $d$ (distance) is the distance a neighbor node has to the destination and $t$ (trust) is the proportion of successfully delivered messages through that node before. When $r$ is computed for every neighbor the message is forwarded to the neighbor with the minimum $r$. When $r$ falls below a certain threshold, for example 0.33, the node considers this neighbor as a void in the distribution space. Backtracking is a standard way developed for geographic routing that allows for efficient routing around voids.

The reputation of a neighbor node can increase by adding a fixed number at every time step until it reaches the threshold level. It is also possible to add trust for every message that was routed through some other node because the geographically optimal node is below the threshold of trust. Message forwarding to border nodes is entirely based on trust since nodes in one cluster are completely oblivious to what nodes are in other clusters and how far away they are.

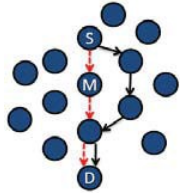
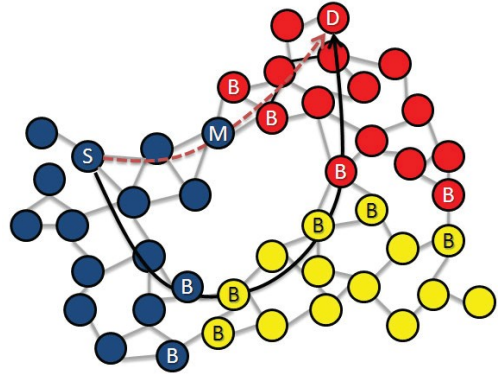

Figure 3: Routing from the source 'S' to the destination 'D' around a malicious node 'M'. Red dashed line is the optimal route, the black solid line is the route actually taken.

Figure 4: Routing between clusters when a border node is malicious.

#### 4.2.1 Inter-Cluster Routing

When messages have to be routed outside a cluster they must be sent to a border node. Thus, each border node must have a way to know if the destination node is reachable through the outside cluster it is connected to. To keep the protocol simple and efficient border nodes do not need to know exactly where the destination is or even its cluster affiliation. This lookup can be implemented very efficiently by using a bloom filter. Bloom filters take very little space and can be queried with simple bit operations. They can be aggregated over a certain number of neighbor clusters. This will guarantee that the node with the destination id is located in one of these neighbor clusters.

If the bloom filter gives a positive result the message is passed to the border node on the other side (in the neighbor cluster) which has to decide if the destination node is in the cluster and should be forwarded internally or the message has to be passed to another border node which will make further decisions. When a source node has to decide to which border node to forward to it consults its cache of border nodes for given destination ids. We focus on caching later in this report. If the node id is not in the cache the source node must flood the border nodes to discover a route through them. Three situations can arise in which the sender node must confirm with the border node the next action. That is, the source node will only send probes on the first round. Then it will react accordingly:

- One border node gives positive response: the node id that has to be located can be found through this border node optimally.

- More than one of the border nodes has the id in the aggregated bloom filter: all of them deliver the package. The ACK to the sender will carry a cumulatively authenticated list of traversed nodes and the sender will now know which of the border nodes is on the shortest path. The id of this border node will be cached as corresponding to a particular id.

- None of the border nodes find the id in their bloom filters. It might be that the id does not exist or is beyond the maximum number of cluster hops for which border nodes keep information. The sender node can issue a request for flooding the border nodes on the whole network. Such requests must be limited. Ideally such situations will occur rarely. Whether flooding is necessary or not depends on how many clusters there are and how much information a border node keeps. The tradeoff here is that the border nodes aggregate a bloom filter from every other cluster within some range. If the range is large this bloom filter will always indicate a match. A search for a node id outside the originating cluster will trigger all border nodes for this cluster and will effectively flood the network a number of times proportional to the number of neighbor clusters. On the other hand, not knowing that the id we are searching for can be reached will also result in a similar message flood.

### 4.2.2 Caching and Cumulative Authentication

Caching is of immense importance for the efficient routing which is outside of the cluster boundaries. Ideally internal nodes will know which border node to contact in order to have the message take an optimal path. Paths outside the cluster are discovered on demand and the initial discovery can, in the worst case, traverse every border node in the system. On the other hand, in practical deployments, it can be expected that nodes would coordinate internally based on geographic proximity and that only a few far-away nodes would need to be contacted. In SAHER we propose that both border nodes and internal nodes cache aggressively. Because border nodes devote all of their resources to routing it can be expected that they have more memory to spare for route lookup tables. Further caches can be on two levels: when a response is received the whole path can be cached and the path reused. Later, as the cache memory is being filled the older entries are reduced to one-hop caches. In a one-hop cache each node stores what the best node to forward to is for each known destination. A whole path cache entry can be submitted with the packet as a suggestion what route the packet should take. This could be ignored or altered by nodes on the path that receive if the network is very dynamic. So, full path caches should be small.

For each package successfully delivered to the destination an ACK is sent. With the ACK a MAC over the list of traversed nodes (cumulative authentication for securing path vector protocols [10]) is sent back all the way to the sender. In this way the path can be cached. Additionally on every hop a local ACK is submitted that has a lifetime of two hops. This allows for building trust among neighbors.

### 4.2.3 Credit Based Alternate Route Enforcement

It is obvious that if the protocol is run as is described above it is very likely that certain nodes will be asked to forward more messages than others because they are reliable and because they stand on a very important route. This will also make them susceptible to flooding and ultimately shutdown due to battery depletion. In SA-HER we prevent this by introducing a credit based system. Each node has a certain amount of credits for every neighbor. Credits recharge with time at a predefined rate. This allows for packet bursts while keeping the amount of traffic that passes through a single node under control. Packets that are received by a neighbor which has exhausted its credits will be dropped. Sender nodes will not be notified why packets are dropped; it is their responsibility to keep track of how many remaining credits there are for each neighbor. On the other hand, if a node drops packets before credits for it have been exhausted it will have its trust diminished and eventually avoided. The introduction of credits promotes load balancing in a distributed way. The amount of credits remaining for a neighbor node can be included in the routing decision formula which can now be rewritten as: $r = \frac{d}{t \cdot c}$ where $c$ is the proportion of remaining credits to the maximum number of credits possible.

### 4.2.4 Join/Leave Operations

It is realistic to expect that nodes can change their location or die unexpectedly.

- Leaving the cluster does not require any notification. Members of a cluster are expected to periodically inform the cluster that they are still part of it. This can be achieved by flooding at predefined intervals.

- Joining a new cluster after it has been formed is slightly more complicated and depends on the way that the cluster size is limited. If it is based on the total number of nodes allowed the procedure for adding the node follows that of new node discovery in the initialization phase. Another approach is to allow a new node to join if its neighbors have fewer neighbors than some threshold. This limits the density of the network at a place. Sending join messages should be rate limited not unlike other messages are limited.

When something in the clusters changes border nodes inform border nodes from neighboring clusters up to the hop count to which they hold bloom filters.

# 5 Results and Discussion

To examine the behavior of the SAHER design we implemented the protocol in a custom simulator. Below we present the simulator and the experimental setup; then we comment on the results and draw some comparisons.

## 5.1 Simulation Environment

To evaluate different aspects of the SAHER protocol, we developed the SAHER simulator. This provides a way to test algorithms at a higher level, without the need for a comprehensive simulation of the underlying operating system and link-layer MAC protocol.

### 5.1.1 Technical Aspects

The simulator is written in ANSI C and uses SDL [28] to implement a graphical frontend. Nodes are modeled as structures and dynamic function pointers are used to enable different behaviors for different types of nodes. Every node only has access to its own data. Communication is handled via a buffer in which all messages are stored for some time. The delivery time is calculated based on the distance between the nodes, to approximate the propagation of the radio waves. Time and energy are counted in artificial units. Time is also counted as discrete cycles. Energy is depleted by transmitting messages where each send operation takes one energy unit. Receiving messages and performing calculations costs no energy in our model. This is in agreement with the fact that most sensor nodes require significantly more energy for transmission than for any other part of their operation. Parallel operation of the individual nodes is simulated by sequentially executing the behavior function of each node for every time unit. During this step a node may read its incoming message buffer or send a message.

The simulator is based on an abstraction of a sensor node. Therefore it assumes a functional MAC-layer for collision free message passing. During initialization nodes are randomly placed inside an area of dimensions passed as arguments to the program. The dimensions of this area along with the transmission range of each node determine the number of neighbors any given node can reach. Off these nodes a specified percentage can be malicious. Malicious nodes can either try to flood the network, drop all ACK packets or drop all packets. The cluster formation can be initialized either randomly or depending on the node id. Cluster size is determined by the hop distance from the initializing node. A node becomes a border node if it has connection to another cluster. After the initialization is complete, nodes start to send payload messages. A receiving node can be either selected randomly or biased by proximity to the sender. Besides the routing table for their own group each node also has a routing cache in which it stores the last $n$ pairs of sender id and the corresponding neighbor. It also uses a message cache to avoid rebroadcasting messages. During runtime the user can insert new nodes into the network or force messages to be send between nodes manually. He can also request information about the current status of each node. This information can also be written to a file at specified intervals during the simulation or after the simulation ends.

### 5.1.2 Experimental Setup

For our experiments we have used the following parameters unless other values were explicitly specified.

- grid size: 200x150
- startup energy: 100,000 units
- cluster hop radius: 8
- maximum number of credits: 30
- credit recharge rate: 1 credit per 30 units of time
- trust recharge rate: 1 point per 100 cycles

## 5.2 Security Analysis

The security of the protocol was examined in the situation where malicious nodes drop packets. The results are illustrated in figure 5. The numbers are averaged over 50,000 cycles in a random node distribution. As expected, there are messages that are lost and have to be resent but even at high concentrations of malicious nodes the number of lost messages is rather low. When there are no malicious nodes ordinary routing takes place and no messages are lost. The number of lost messages reaches a maximum of around 8% when the malicious nodes make up about half of the network. We believe that the number of dropped messages becomes lower at higher concentrations of dropping nodes because there are clusters that only have adversaries as border nodes. These clusters can be efficiently routed around.

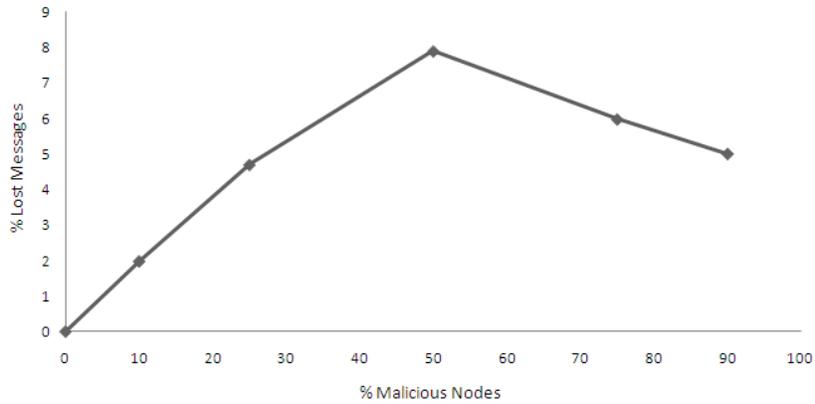

Figure 5: Percentage of lost messages with different concentrations of malicious nodes in the network.

For comparison, first consider the simplest type of protocol broadcast. In a broadcast protocol there would be no lost messages if at least one path without malicious nodes exists that connects the source and the destination. This however will result in enormous resource consumption as each message will trigger all nodes to transmit. Also, issues with collisions and concurrency on lower levels can occur. The other extreme is geographic routing where a path that is very close to the optimal will be found greedily. This protocol is probably the most efficient when there are no attacks on the network layer. However, since the protocol is greedy even with a few nodes that drop every packet some parts of the network will become disconnected as nodes will blindly forward messages to malicious neighbors. There are some very interesting recent examples of secure routing protocols that exploit similar ideas. An example is the SCR [4] protocol which bears some similarity to SAHER in that the protocol is based on a clustered architecture. This protocol however relies on at least one base station to function correctly. The rate of dropped packets for SCR can exceed 30% if a malicious node on a path drops packets with probability 90%. Tanachaiwiwat et al. introduce another protocol [27] that implements security by exploiting the idea of secure geographic routing built on trust, similar to what is proposed here. However there is no cluster-level isolation thus memory requirements for this protocol will be very high for large networks. In terms of delivery ratio the results are close to what we observe on SAHER.

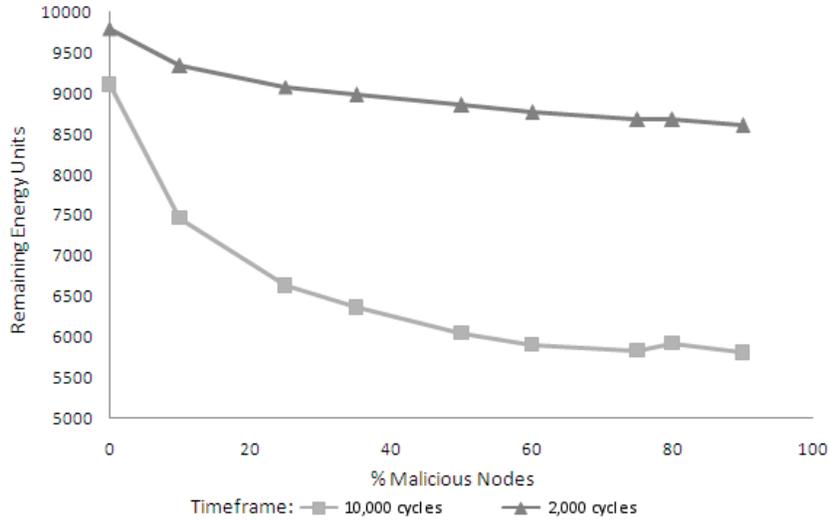

Figure 6: Remaining energy in alive fair nodes.

In section 4.2.3 we added a credit based system to the protocol to alleviate the effects of packet flooding. All packets, including join/leave packets, are limited in the same way. In figure 6 we present the remaining energy units for different concentrations of malicious nodes and different time frames. The numbers presented are the average number of energy units for fair nodes that have not depleted their energy yet. It can be clearly seen that credits provide a very good protection against flooding. The amount of damage that even 90% of malicous nodes can deal is very limited.

## 5.3 Efficiency

Because minimizing resource consumption is just as important as ensuring security we examine the memory requirements, computational requirements and the overhead messages.

SAHER has minimal extra memory requirements. Internal nodes need to keep a table of coordinates for nodes within their cluster. We can assume that ids are 16 bit long allowing $2^{16}$ nodes in the network. Global coordinates can also be held in 16 bit. Thus, a cluster size of 100 nodes requires a table of 3.2KB. Border nodes only need to keep an additional bloom filter. If their knowledge spans two hops for the same cluster, the bloom filter will only need 2KB with 1% chance of false positives for the same cluster size. The rest of the memory can be used for cache, operating system and application programs. CPU requirements are just as low since bloom filter query and table lookup are comprised of a few simple operations.

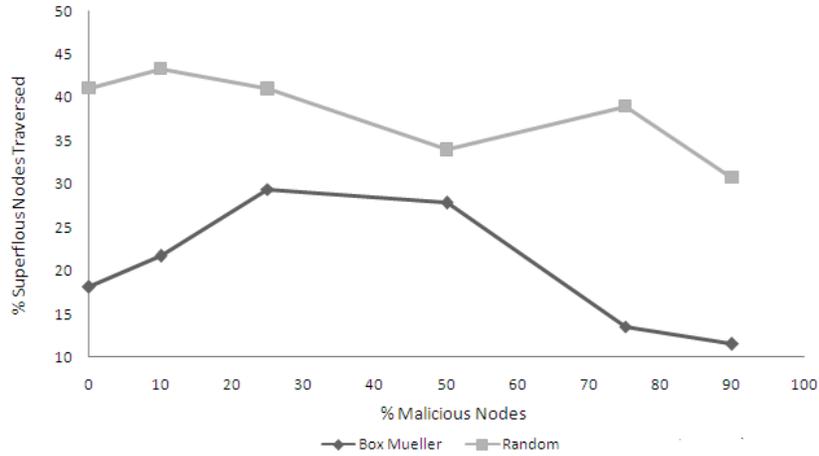

Figure 7: Suboptimal hops.

An important measurement of protocol efficiency is the message overhead it introduces. In figure 7 we present the number of superfluous nodes that are traversed on a path. The percentage of route length increase is depicted. That is, if the optimal path is 10 hops and the path taken is 11 hops the percentage of unnecessary nodes traversed is 10%. We present two scenarios: one with a completely random destination and source; and another with a destination node selected with an increased probability from the neighborhood of the source. For the second scenario we use normal distribution emulated with the Box-Mueller algorithm. With a random source and destination selection the maximum hop overhead is 40%. As expected, if destination nodes are selected to be close to the source the overhead is much lower with a peak at 30%. It is interesting to note that as the number of malicious nodes increases there is a trend that the overhead is lower because only few routes without malicious nodes remain and all messages are passed through them.

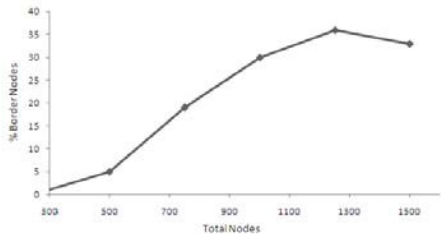
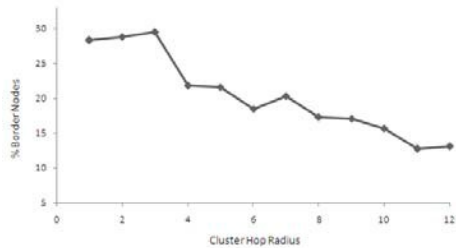

Figure 8: The proportion of border nodes increases and reaches a saturation point as the size of the network increases.

Figure 9: Creating larger clusters reduces the number of border nodes.

SAHER requires border nodes to relinquish normal operation and focus ex-clusively on supporting the communication between clusters so it is interesting to examine the proportion of border nodes to normal nodes in different settings for the simple clustering and border node selection that we simulate. This is presented in figures 8 and 9. In figure 10 the tradeoff on communication during setup and the cluster size is illustrated for startup energy of 50,000 units, where group formation takes around 250 cycles. In summary, larger groups naturally take more energy to self-organize but the network needs less border nodes.

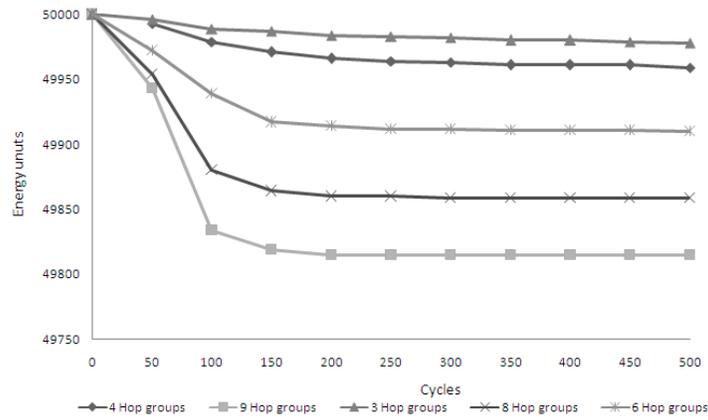

Figure 10: Energy used in cluster formation.

# 6 Conclusion and Future Work

In this report we introduced a clustered architecture around which we build SAHER, a secure routing protocol for sensor networks. Due to the fact that sensor nodes possess extremely limited computational and power resources we also considered efficiency as a primary objective. Our evaluation shows that the protocol is resilient against malicious nodes that drop packets by letting each node maintain a trust value for its neighbors. This forms the basis for alternative route selection, triggered when malicious nodes happen to be a part of the optimal path. Also, flooding can be prevented in a similar way where nodes have message credits with their neighbors; when that is exceeded alternative routes have to be taken. Only efficient primitives are used: geographic routing, bloom filters and symmetric cryptography ensure low resource consumption. As in other protocols adding security introduces some overhead. We incur some overhead in the cluster formation, routing messages to nodes outside the source clusters and redundant ACKs. This overhead however is not excessive and is a relatively small price to pay for the benefits of increased robustness.

There are some issues that could be addressed in the future. More experimentation has to be done on the clustering algorithms and parameters for optimal cluster formation. Perhaps an entirely new clustering scheme has to be designed in order to avoid the large amount of resources consumed in the initialization stage. The trust model is rather simplistic and further research could focus on making it better. Also, other measures of performance and security with different setup parameters could be included. A general improvement might be to extend the node model in the simulator to more closely follow a specific sensor node design and some real-world applications. This could allow us to closely emulate the behavior of nodes in potential deployments.